# Ferromagnetic MnBi₄Te₇ obtained with low concentration Sb doping: a promising platform for exploring topological quantum states

Y.D. Guan[1+], C.H. Yan[2+], S.H. Lee[1,3], X. Gui[4,] W. Ning[1], J.L. Ning[5], Y.L. Zhu[1,3], M. Kothakonda[5], C.Q. Xu[6], X.L. Ke[6], J.W. Sun[5], W.W. Xie[7], S.L. Yang[2*] and Z.Q. Mao[1,3,8*]

[1] Department of Physics, The Pennsylvania State University, University Park, PA 16802, USA

[2] Pritzker School of Molecular Engineering, The University of Chicago, Chicago, Illinois 60637, USA

[3] 2D Crystal Consortium, Materials Research Institute, The Pennsylvania State University, University Park, PA 16802, USA

[4] Department of Chemistry, Louisiana State University, Baton Rouge LA 70803

[5] Department of Physics and Engineering Physics, Tulane University, New Orleans, LA 70118

[6] Department of Physics and Astronomy, Michigan State University, East Lansing, MI 48824, USA

[7] Department of Chemistry and Chemical Biology, Rutgers University, Piscataway NJ 08854

[8] Department of Materials Science and Engineering, The Pennsylvania State University, University Park, PA 16802, USA

## Abstract

The tuning of magnetic phase, chemical potential, and structure is crucial to observe diverse exotic topological quantum states in $MnBi_2Te_4(Bi_2Te_3)_m$ ($m = 0, 1, 2, \& 3$). Here we show a ferromagnetic (FM) phase with a chiral crystal structure in $Mn(Bi_{1-x}Sb_x)_4Te_7$, obtained via tuning the growth conditions and Sb concentration. Unlike previously reported $Mn(Bi_{1-x}Sb_x)_4Te_7$, which exhibits FM transitions only at high Sb doping levels, our samples show FM transitions ($T_C = 13.5$ K) at 15%-27% doping levels. Furthermore, our single crystal x-ray diffraction structure refinements find Sb doping leads to a chiral structure with the space group of $P3$, contrasted with the centrosymmetric $P\bar{3}m1$ crystal structure of the parent compound MnBi$_4$Te$_7$. Through ARPES measurements, we also demonstrated that the non-trivial band topology is preserved in the Sb-doped FM samples. Given that the non-trivial band topology of this system remains robust for low Sb doping levels, our success in making FM Mn(Bi$_{1-x}$Sb$_x$)$_4$Te$_7$ with $x$ = 0.15, 0.175, 0.2 & 0.27 paves the way for realizing the predicted topological quantum states such as axion insulator and Weyl semimetals. Additionally, we also observed magnetic glassy behavior in both antiferromagnetic MnBi$_4$Te$_7$ and FM Mn(Bi$_{1-x}$Sb$_x$)$_4$Te$_7$ samples, which we believe originates from cluster spin glass phases coexisting with long-range AFM/FM orders. We have also discussed how the antisite Mn ions impact the interlayer magnetic coupling and how FM interlayer coupling is stabilized in this system.

+Y.G. Guan & C.H. Yan have equal contributions to this work

*Email: yangsl@uchicago.edu; zim1@psu.edu

## I. INTRODUCTION

Magnetic topological quantum materials have attracted considerable attention due to its great potential in generating emergent topological quantum phenomena of technological relevance, such as quantum anomalous Hall effect (QAHE) and quantum axion electrodynamics [1-3]. Since QAHE is accompanied by the chiral edge channels, which can support dispersionless current, it holds great promise for potential applications in novel spintronics and topotronics. QAHE was first demonstrated in the topological insulator (TI) thin films of Cr- and/or V-doped $(Bi,Sb)_2Te_3$ [4,5]. However, due to the doping-induced chemical inhomogeneity, the required temperatures for observing QAHE is limited to a sub-kelvin range. Realization of QAHE at higher temperatures demands a combination of non-trivial band topology and intrinsic magnetic orders.

$MnBi_2Te_4$ is the first established intrinsic antiferromagnetic (AFM) TI [6-8] and has been found to host a variety of exotic quantum states. When its crystal thickness is reduced to a few layers, it shows either QAHE [9,10] or an axion insulator behavior [11,12], depending on the number of layers. Moreover, it also exhibits Chern insulator behavior in both ferromagnetic (FM) and canted AFM states [13-15]. For bulk single crystals, when its magnetic order is driven from AFM to FM by an external magnetic field along the $c$-axis, its electronic state transits from a TI to a type-II ideal Weyl semimetal with only one pair of Weyl nodes [6,7,16]. These diverse topological quantum states seen in $MnBi_2Te_4$ have generated enormous interests. Additionally, recent theoretic studies predict this material can also show other exotic quantum phenomena under certain situations, such as high order TI [17], giant linearly-polarized photogalvanic effect and second harmonic generation [18], Majorana hinge mode [19], magnetic Skyrmion lattice [20], etc.

The fascinating exotic properties seen/predicted in $MnBi_2Te_4$ have motivated studies on other related materials, including $MnBi_4Te_7$, $MnBi_6Te_{10}$, and $MnBi_8Te_{13}$. These materials belong to the same family, which can be expressed as $(MnBi_2Te_4)(Bi_2Te_3)_m$ with $m = 1, 2, 3, \ldots$ Their common structural characteristic is the alternating stacking of $[MnBi_2Te_4]$-septuple layers (SLs) and $[Bi_2Te_3]$-quintuple layers (QLs). The main difference between the $m = 1, 2, 3$ members is the number of QLs ($m$) sandwiched between SLs; $m = 1$ for $MnBi_4Te_7$, $m = 2$ for $MnBi_6Te_{10}$, and $m = 3$ for $MnBi_8Te_{13}$. These materials show tunable band topology and magnetism. The increase of the number of QLs weakens the interlayer AFM coupling, leading the system to evolve from AFM to FM states [7,21-24]. Theory predicts the combination of non-trivial band topology and tunable magnetism in $(MnBi_2Te_4)(Bi_2Te_3)_m$ can generate a rich variety of topological quantum states [21,22,24-27] such as AFM TI [22], axion insulator [26,27], FM Weyl semimetals [28,29], QAH insulator [30], Möbius insulator [17], high Chern number quantum spin Hall insulator [31], etc. Experimentally, $MnBi_4Te_7$ has been shown to be an AFM TI [32], $MnBi_6Te_{10}$ is either an AFM TI [26,33] or an AFM axion insulator [27], whereas $MnBi_8Te_{13}$ is a FM axion insulator [34]. $MnBi_4Te_7$ and $MnBi_6Te_{10}$ may also become axion insulators or Weyl semimetals if their magnetic orders become FM according to theoretical studies [28,29,35]. Although first principle calculations show their FM phases have a slightly higher total energy than the AFM phases ($\Delta E = E_{FM} - E_{AFM} = 0.23$ meV for $MnBi_4Te_7$, 0.05 meV for $MnBi_6Te_{10}$ [36]), the synthesis of FM phases is challenging for these materials. Nevertheless, their FM phases are found to be accessible by Sb doping to the Bi site in $Mn(Bi_{1-}$ $_x$Sb$_x$)$_4$Te$_7$ [29,37]. Hu *et al*. [29] reported that the system exhibits an AFM transition at $T_N = 13$ K, then followed by a FM-like transition at a lower temperature $T_C$ for $0 < x < 0.6$, but only a FM transition for $x > 0.6$. However, another earlier work by Chen *et al*. [37] shows inconsistent result: while most of their Mn(Bi$_{1-x}$Sb$_x$)$_4$Te$_7$ samples with various values of $x$ (=0.15, 0.36, 0.4 & 0.46) show AFM-to-FM transitions similar to those reported by Hu *et al*. [29], their $x = 0.3$ sample was found to show only a FM transition at about 13K. These inconsistent results imply the formation of FM phase in MnBi$_4$Te$_7$ is dependent on not only Sb concentration, but also the level of disorders controlled by synthesis conditions. Gaining a systematic control of the FM phase in Mn(Bi$_{1-x}$Sb$_x$)$_4$Te$_7$ is crucial to pursue those topological quantum states observable only in spontaneous FM phases, but is still lacking at present.

In this article, we will show the FM phase of Mn(Bi$_{1-x}$Sb$_x$)$_4$Te$_7$ can be reproducibly synthesized with 15-27% Sb doping via tuning growth conditions, and demonstrate the non-trivial band topology is preserved in our synthesized FM samples through ARPES measurements. Moreover, we have also observed magnetic glassy behavior for both pristine AFM and Sb-doped FM samples, which can be attributed to the coexistence of the cluster spin glass phase (CSG) and long-range AFM/FM order. Additionally, we find the Sb doping lowers crystal symmetry, resulting in a chiral structure with the space group of *P3*. These findings establish a new material platform which combines non-trivial band topology, spontaneous ferromagnetism, and chiral structure. Such a material provides a promising platform to pursue those theoretically predicted topological quantum states noted above, such as axion and Weyl semimetals.

## II.    METHODS

Mn(Bi$_{1-x}$Sb$_x$)$_4$Te$_7$ single crystals were grown using a melt growth method. High purity powders of Mn, Bi, Sb, and Te were mixed with molar ratios of Mn: Bi: Sb: Te= 1: 4-4$x$: 4$x$: 7 ($x$ = 0, 0.15, 0.175, 0.2, and 0.27). The mixtures of source materials were loaded into quartz tubes and sealed under high vacuum. The inner walls of the quartz tubes were coated with carbon to prevent the reaction between Mn and silica ampoules. The quartz tubes loaded with source materials were heated in muffle furnaces up to 900 °C and dwelled at this temperature for 5 hours to ensure homogeneous melting. Then the mixtures were quickly cooled down to 595 °C within 20 hours, followed by slowly cooling down to 585 °C in three days (~ 0.14 °C/h). Then the samples were annealed at 585 °C for another day and finally quenched in water. For some MnBi$_4$Te$_7$ samples, we also performed additional heat treatments with the aim of introducing more disorders. Right after one-day annealing at 585 °C, we first decreased the temperature down to 400 °C and then increased the temperature back to 585 °C before water quenching. This additional step is expected to increase the anti-site defects according to our previous studies on Mn(Bi$_{1-x}$Sb$_x$)$_2$Te$_4$ [38]. Hereafter, we will label the pristine MnBi$_4$Te$_7$ sample without undergoing additional heat treatment as MBT1 and the sample with additional heat treatment as MBT2. Mn(Bi$_{1-x}$Sb$_x$)$_4$Te$_7$ samples with $x$ = 0.15, 0.175, 0.2, and 0.27 were synthesized following a similar procedure as the one used for synthesis of sample MBT1 and are denoted by MBST1 (15% Sb), MBST2(17.5% Sb), MBST3 (20% Sb) and MBST4 (27% Sb) respectively below.

All samples selected for magnetic, transport, and ARPES measurements were characterized using X-ray diffraction (XRD) and confirmed to have the desired phase. For the 15% Sb doped sample, we also carried out single-crystal X-ray diffraction (SCXRD) structure refinement to find how Sb doping alters the structure. The composition analyses using X-ray energy dispersive spectroscopy (EDS) measurements show that the actual Sb/Bi ratio probed in the Sb-doped samples deviates from the nominal ratio only by < 2.5%. EDS mapping also does not show any clusters of Bi/Sb, suggesting relative homogeneity of the Sb-doped samples. Magnetic properties of the synthesized materials were measured using a Superconducting Quantum Interference Device (SQUID, Quantum Design). Magnetotransport measurements were conducted using a four-probe method in a Physical Property Measurement System (PPMS, Quantum Design). ARPES measurements were conducted using the ultrahigh resolution 6 eV ARPES module on the platform of multi-resolution photoemission spectroscopy (MRPES) established at the University of Chicago [39]. The overall energy resolution and angular resolution were 4 meV and 0.3°, respectively. The laser beam waist was ~10 $\mu m$.

## III. RESULTS

Figure 1c presents representative (00$l$) XRD patterns of samples MBT1, MBT2, MBST1, MBST3 and MBST4. Sharp diffraction peaks seen in these patterns indicate high crystallinity quality of these samples. Pristine $MnBi_4Te_7$ possesses a layered trigonal structure with the space group of $P\bar{3}m1$ [22], which can be viewed as a naturally formed superlattice, composed of alternating stacking of the [$MnBi_2Te_4$]-SLs and [$Bi_2Te_3$]-QLs, as shown in Fig. 1a. However, our SCXRD structure refinement finds the Sb doping leads to a substantial structural change. In contrast with the centrosymmetric trigonal structure of $MnBi_4Te_7$, the 15% Sb-doped sample (MBST1) crystallizes into a chiral trigonal structure with the space group of $P3$, as discussed below. Our refinement results show such a chiral structure arises from preferential Sb occupations at the Bi site. While Sb is mixed with Bi on layers 1, 2, and 4 shown in Fig. 1b, with the Sb occupation rate being ~ 30% for layers 1 & 2 and ~ 70% for layers 4, there are hardly Sb occupations on layers 3.

We performed SCXRD measurements on several pieces of MBST1 to verify the refined chiral structure. The XRD data were collected at $T = 296$ K and Fig. 1d present representative Bragg diffraction patterns on three crystallographic planes. According to Friedel's Law, the atomic scattering factors (X-ray scattering intensities) associated with two reciprocal points at (***h, k, l***) and (***-h, -k, -l***) are almost equal. Thus, it is impossible to determine the chirality from the powder X-ray patterns. However, atomic scattering factors have imaginary parts due to the anomalous dispersion effect in chiral structures. To confirm the existence of chirality, we focused on seeking the anomalous scattering in the crystal, which should result in a small difference in scattering intensity. Such a small intensity difference can be detected directly from single-crystal X-ray diffraction. In Figs. 1e - 1f, we present the X-ray scattering intensity maps of the crystal in the reciprocal space, projected along the c*- axis (Fig. 1f) and a zone-axis slightly away from c* (Fig. 1e). Notably, the scattering intensity distribution in Fig. 1e is non-symmetric, which is a typical characteristic of a crystal structure without inversion symmetry. This indicates the measured single crystal should possess a non-centrosymmetric crystal structure. With 7993 diffraction peaks collected in our SCXRD measurements (see Table 1), we performed structure refinements, which yields the best refined structure of *P3* as noted above. The inversion symmetry breaking likely leads the structure with preferential Sb occupations to be more stable. Although the DFT calculations cannot resolve the total energy difference of the ground state between uniform and preferential Sb occupations (see Supplementary Note 1 and Table S1 [40]), lattice dynamics at finite temperature might be different for the uniform and preferential Sb occupations; preferential Sb occupations may have lower free energies at high temperatures in comparison with the uniform Sb occupations (Note that like $MnBi_2Te_4$, $MnBi_4Te_7$ is also a metastable phase and can be synthesized only through quenching at high temperatures). Further understanding of preferential Sb occupations requires more detailed investigations.

The composition obtained from the structure refinements is $(Mn_{0.80(1)}Bi_{0.20(1)})(Bi_{2.71(4)}Sb_{1.29(4)})Te_7$, suggesting ~20% Mn sites are occupied by (Bi,Sb), consistent with previously reported compositions obtained from X-ray and neutron diffraction refinements [28,41,42]. Note that due to the refinement limitation of elements, we can only mix two types of atoms on one atomic site. It is assumed that only Mn and Bi atoms mix on the Mn site and Sb and Bi atoms mix on the Bi site. The atomic occupancy on Te sites was tested, and no vacancies was detected. The Bi-Sb mixing at the Mn site and antisite Mn occupation at the Bi/Sb site were not taken into account in our structure refinements though these scenarios should exist. The structure parameters obtained from the refinements as well as some other refinement information are summarized in Table 1.

We have studied the magnetic properties of samples MBT1, MBT2, and MBST1-4 through magnetization measurements. The temperature dependences of magnetic susceptibility $\chi(T)$ measured with $H \perp ab$ for sample MBT1 and MBT2 are presented in Fig. 2a and 2d, respectively. Although both samples show AFM transitions at $T_N = 13$ K, they display distinct irreversible behaviors. Sample MBT1 exhibits bifurcation between zero-field-cooling (ZFC) and field cooling (FC) below $T_{ir} \sim 8$ K, consistent with prior reports [28]. However, in sample MBT2, we find the susceptibility irreversible behavior starts to emerge from ~10.5 K and becomes much more significant than that in sample MBT1 below 5K, implying that the additional heat treatments experienced by sample MBT2 alters its magnetic properties. To reveal the essential difference in magnetic properties between these two samples, we have measured their isothermal magnetizations $M(H)$ at various temperatures. The data obtained from these measurements are presented in Fig. 2b-2c for MBT1 and Fig. 2e-2f for MBT2, which clearly show both MBT1 and MBT2 samples undergo spin-flip/flop transitions upon the out-of-plane field sweeps for both $T_{ir} < T < T_N$ (Fig. 2c & 2f) and $T < T_{ir}$ (Fig. 2b & 2e) except for 2 K. Both samples exhibit FM-like polarization behavior at 2 K, similar to the result previously reported by Vidal *et al*. [28].

Furthermore, we have also measured $\chi(T)$ and $M(H)$ with $H//ab$ for samples MBT1 and MBT2. The data obtained from these measurements are presented in supplementary Figs. S1(a,b) and S2(a,b) [40], which reveal strong magnetic anisotropy in both samples. Their $\chi(T)$ measured with $H//ab$ shows upturn below $T_N$, contrasted with the sharp peak of $\chi(T)$ observed under $H\perp ab$. Unlike $M(H)$ measured under $H\perp ab$ which displays step-like spin flop/flip transitions at low fields (<0.15 T), $M(H)$ measured under $H//ab$ shows gradual spin flop/flip transitions and its saturation field is about 10 times larger than that of $H\perp ab$, consistent with prior reports [21,23,28,32].The results described above suggest the magnetic states of both MBT1 and MBT2 samples are characterized by out-of-plane AFM orders at temperatures below $T_N$ and above 2 K, but possibly evolve into out-of-plane FM orders below 2 K. Nevertheless, our magnetotransport measurements presented below suggest that the magnetic ground state is indeed characterized by the coexistence of FM and AFM phases.

Previous studies on $Mn(Bi_{1-x}Sb_x)_2Te_4$ have shown Mn antisite defects favor interlayer FM

coupling and can be increased significantly with Sb substitution for Bi [38,43,44]. A ferrimagnetic order mediated by antisite Mn has been demonstrated in $MnSb_2Te_4$ [43,44] and $Mn(Sb_{1.8}Bi_{0.2})Te_4$ [38]. Such a ferrimagnetic order results in interlayer FM coupling, which leads the materials to show FM-like transitions. The ferromagnetic behavior observed by Hu $et\ al.$ [29] in $Mn(Bi_{1-x}Sb_x)_4Te_7$ with $x > 0.6$ follows a similar mechanism. Using our synthesis conditions described above, we have synthesized FM $Mn(Bi_{1-x}Sb_x)_4Te_7$ samples with much lower Sb doping concentrations ($x = 0.15$, 0.175, 0.20 and 0.27). The magnetic susceptibility data measured with $H \perp ab$ for samples MBST1, MBST3 and MBST4 are shown in Fig. 3a, 3d and 3g, respectively (the data of sample MBST2 is presented in supplementary Fig. S4 [40]). In contrast with samples MBT1 and MBT2 whose susceptibilities show typical characteristics of an AFM transition, i.e., sharp cusps at $T_N = 13$ K (Fig. 2a and 2d), samples MBST1-4 exhibit characteristics of FM transitions at $T_C = 13.5$ K (or 13 K). The susceptibility bifurcation between FC and ZFC occurs when the temperature is decreased below $T_{ir} = 9$ K (or 8 K) in all these samples. Unlike conventional ferromagnets whose magnetic susceptibility under a FC history usually tends to saturate at temperatures well below $T_C$, samples MBST1-4 show steep upturns in the susceptibilities measured with FC histories below $T_{ir}$, consistent with the prior report by Hu $et\ al.$ [29]. To further understand this unusual feature, we have also measured in-plane resistivity for these samples as well as specific heat for sample MBST1. The data from these measurements have been added to Fig. 3a, 3d and 3g. The FM transitions at $T_C$ are manifested by resistivity reduction below $T_C$ and an anomalous specific heat peak at $T_C$. However, at $T_{ir}$ where the FC susceptibility upturn occurs, both the resistivity and specific heat do not show any anomalies, which excludes the possibility that a secondary magnetic phase transition occurs at $T_{ir}$. As will be discussed below, this feature as well as the magnetic irreversible behavior observed in these samples indicate their FM states have glassy nature below $T_{ir}$.

The FM transitions described above for samples MBST1-4 are also evidenced by isothermal magnetization measurements, as shown in Figs. 3b, 3e, and 3h, respectively. Unlike samples MBT1 and MBT2 which show field-driven spin-flip/flop transitions, samples MBST1-4 do not show any spin-flip/flop transitions, but FM domain polarization behavior upon magnetic field sweeping for both $T_{ir} < T < T_N$ (Fig. 3c, 3f & 3i) and $T < T_{ir}$ (Fig. 3b, 3e &3h). Like samples MBT1 and MBT2, samples MBST1-4 also exhibit magnetic hysteresis only for $T < T_{ir}$. Their magnetic coercive force $H_c$ is small, < 200 Oe at 3 K, indicating that they are very soft FM materials. These results, together with the $\chi(T)$ and $M(H)$ data measured with $H//ab$ for MBST1-4 (see supplementary Fig. S3(a,b) for MBST1, Fig. S4(a,b) for MBST2, Fig. S5(a,b) for MBST3, and Fig. S6(a,b) for MBST4 [40]) indicate all the MBST1-4 samples exhibit out-of-plane ferromagnetism below $T_C$. To show the reproducibility of both antiferromagnetism and ferromagnetism in our MBT and MBST samples, we measured multiple pieces of samples for each MBT/MBST batches and added the additional data of $\chi(T)$ and $M(H)$ obtained from these measurements to Supplementary Figs. S1-6 [40]. While all the measured pieces have similar $T_N$ or $T_C$, the measured saturation moment $M_s$ varies between different pieces of samples even for the samples from the same batch; for instance, $M_s$ varies in the 2.55 -3.05 $\mu_B$/f.u. range for the MBST3 batch. In Table 2, we list the variation range of $M_S$ for every MBST/MBT batch as well as the averaged value $\bar{M}_S$ (i.e. the values shown in the parentheses). $\bar{M}_S$ of the MBST samples is ~ 3%-10% less than that of MBT1, the origin of which will be discussed below.

Although all the samples studied here, including MBT1-2 and MBST1-4, show long-range magnetic ordering below $T_N$ or $T_C$ as discussed above (AFM for MBT1 & MBT2; FM for MBST1-4), they all display magnetic glassy behavior in the $T < T_{ir}$ temperature regimes where the susceptibility bifurcation between ZFC and FC occurs. This is revealed by magnetic relaxation measurements shown in Fig. 4a-4c where time dependences of magnetization at various temperatures are presented for samples MBT1, MBT2 and MBST1. Each measurement shown there was conducted right after the magnetic field was ramped up to 0.5 T and then back to zero. The data collected from these measurements are presented with $\Delta M = M(t) - M_0$ on a logarithmic time scale in Fig. 4a-4c; $M_0$ represents the magnetization at 5,000 s (2500 s for MBT1), the time when the measurement ends. From these data, we can see remarkable slow magnetic relaxation for $T < T_{ir}$, which is generally expected for a magnetic state with glassy behavior [45]. FM samples MBST2-4 also exhibit similar magnetic relaxation behavior. The data of MBST3 is presented in supplementary Fig. S7 [40]. For samples MBST1 and MBST3 & 4, we have also conducted AC susceptibility $\chi_{ac}$ measurements. As seen in Fig. 4d-4f, $\chi_{ac}$ shows frequency-dependence below the peak temperature, which is a typical characteristic of a spin-glass or cluster spin-glass state. These results, together with the magnetic irreversible behavior discussed above, indicate all our samples, while showing long-range AFM/FM transitions, are characterized with magnetic glassy behavior below $T_{ir}$. We note slow magnetic relaxation behavior as well as frequency dependent $\chi_{ac}$ have been observed in several related compounds with FM transitions, including Sb-doped $Mn(Bi_{1-x}Sb_x)_6Te_{10}$ ($x = 0.3$ & 0.07) [46], $MnSb_2Te_4$ [47], $Mn(Bi_{0.24}Sb_{0.76})_4Te_7$ [29], $MnBi_8Te_{13}$ [34] and Sb-doped $MnBi_8Te_{13}$[46]. While spin glass is proposed to be the origin of magnetic relaxation in FM MnSb$_2$Te$_4$ [47], this mechanism cannot account for the magnetic relaxation behavior seen in the FM phases obtained in Sb-doped MnBi$_4$Te$_7$ and MnBi$_6$Te$_{10}$ and MnBi$_8$Te$_{13}$[29,37,46]. Single-layer magnet [23] and irreversible domain wall movement [46] have also been proposed to be the magnetic relaxation origins of these compounds. Our observation of magnetic relaxation in both AFM MnBi$_4$Te$_7$ and FM Mn(Bi$_{1-x}$Sb$_x$)$_4$Te$_7$ materials suggests cluster spin glass coexists with long-range AFM/FM order below $T_{ir}$, which we will discuss in detail later.

In addition to magnetic measurements, we have also performed Hall resistivity $\rho_{xy}$ and in-plane longitudinal magnetoresistivity [defined as MR = $(\rho_{xx}(H) - \rho_{xx}(0))/\rho_{xx}(0)$ ] measurements for samples MBT1 (Fig. 5a-5c), MBT2 (Fig. 5d-5f), MBST1 (Fig. 6a-6c), MBST3 (Fig. 6d-6f) and MBST4 (Fig. 6g-6i). The transport signatures revealed through these measurements are mostly consistent with the magnetic transitions revealed by the magnetization measurements. In the AFM samples MBT1&2, we observed anomalous Hall effect (AHE) caused by the spin-flip/flop driven ferromagnetism and the striking magnetic hysteresis below $T_{ir}$ is also manifested by the hysteric behavior between the upward and downward field sweep cycles of $\rho_{xy}$ (Fig. 5b & 5e) and MR (Fig. 5c &5f). It is interesting to note that the field sweeps of MR generate a butterfly-like MR curve at 2.0 K (or 2.5 K) in both samples (Fig. 5c & 5f), consistent with a prior report [23,32,37]. Given the magnetization of samples MBT1&2 exhibits FM-like polarization at 2 K as noted above (Fig. 2b & 2e), their butterfly-like MR curves seen at 2.0 K (or 2.5 K) can be understood as follows: their magnetic states at 2.0 K (or 2.5 K) feature coexistence of AFM and FM phases and the AFM state below the spin-flip/flop transition field should have a positive MR, while the FM state above the spin-flip/flop transition field should have a negative MR due to suppressed spin scattering. The competition of negative and positive MR near the spin-flip/flop transition field leads to a peak in MR. In this case, it is not surprising to see a butterfly-like MR curve when considering strong magnetic hysteresis below 5 K. This interpretation is verified by MR measurements at 2.5 K on the FM samples MBST1-4 (see Fig. 6c, 6f, 6i; the MR and $\rho_{xy}$ data of MBST2 are shown in supplementary Fig. S8 [40]). Since spontaneous ferromagnetism is present below 13.5 K or 13 K in MBST1-4, their MR is dominated by the negative term, which should not generate butterfly-like MR curve upon field upward and downward sweeps. This is exactly what we observed in experiments: all the MBST samples primarily exhibit negative MR, with the positive MR due to the orbital effect being negligible (Fig. 6c, 6f & 6i). The above discussions suggest that while the magnetization data of samples MBT1-2 implies a crossover magnetic transition from an AFM to an FM state below 3 K (Fig. 2b & 2e), coexistence of the AFM and FM phase should occur below 3 K and that pure FM phases can be stabilized by low-concentration Sb doping under our synthesis conditions.

The access to the FM phase of Sb-doped $MnBi_4Te_7$ provides an opportunity to explore new exotic electronic states, since ferromagnetism breaks time-reversal symmetry and causes band splitting. Earlier theoretical calculations indeed suggested the FM $MnBi_4Te_7$ hosts a Weyl semimetal state with multiple pairs of Weyl nodes [28,29]. While we did not probe clear transport signatures of a Weyl state (e.g. chiral anomaly) in our current magnetotransport measurements on the FM MBST1-4 samples, we did find evidence that the electronic structure of the FM MBST samples differs from those of the AFM MBT samples. As seen in Fig. 5a and 5d, the normal Hall contribution due to the Lorentz effect in samples MBT1&2 exhibits linear field dependence, consistent with a prior report [23,28,32]. This indicates the transport properties of the pristine compound are dominated by a single electron pocket though prior studies suggest multiple electron pockets exist in MnBi$_4$Te$_7$ [23,28,32]. However, for the FM MBST1-4 samples, their normal Hall contributions, while also showing the carrier type of electron, display strong deviations from the linear field dependence (see Figs. 6a, 6d, 6g, and Fig. S8a [40]), suggesting a dominant multiple band effect. Such a contrast in the normal Hall contributions between the AFM MBT and FM MBST samples indicates Sb substitution for Bi not only stabilizes the ferromagnetism under our synthesis conditions but also leads to the changes of the band structure near the Fermi level in the FM phase.

To gain more information on the evolution of electronic structure with Sb doping, we estimated the carrier density and mobility of all the MBT and MBST samples using their $\rho_{xy}$ and longitudinal resistivity $\rho_{xx}$ data measured at 200 K where $\rho_{xy}$ exhibits linear field dependence for all the samples as shown in supplementary Fig. S9 [40] (Note that since the MBST1-4 samples exhibit multi-band effects in $\rho_{xy}$ below $T_C$, we chose to compare the carrier density and mobility of the MBT and MBST samples in their paramagnetic states). As shown in Table 2, the carrier density of the MBST1-4 samples is in the $0.28$-$0.87 \times 10^{20}$ cm$^{-3}$ range, one order of magnitude less than that of the MBT1&2 samples. However, the carrier mobility of the MBST1-4 samples is one order of magnitude higher than those of MBT1&2 samples. These variations of carrier density and mobility with Sb doping can be attributed to the chemical potential shift caused by Sb doping as seen in Mn(Bi$_{1-x}$Sb$_x$)$_2$Te$_4$ [16,48-50]. From the Hall resistivity measured at temperatures below $T_C$ (or $T_N$) (Fig. 5b&5e, Fig. 6b, 6e & 6h), we have also obtained anomalous Hall resistivity $\Delta\rho_{xy}$ after subtracting the normal Hall contribution. In supplementary Fig. S10 [40], we also present more Hall resistivity data measured on additional MBT and MBST samples. Table 2 lists $\Delta\rho_{xy}$ for all the samples we measured, including the samples shown in Fig. S10 [40]. We find most MBST samples have comparable $\Delta\rho_{xy}$ with MBT1&2 samples; only one MBST1 sample exhibits relatively larger $\Delta\rho_{xy}$. These anomalous Hall effects observed in the MBST1-4 samples may not directly be associated with the predicted Weyl state, since we did not observe any signatures related to the chiral anomaly expected for a Weyl state as noted above. In supplementary Fig. S11 [40], we present representative data of in-plane magnetoresistivity measured under different field orientations on MBST3 & MBST4 samples, from which we did not observe negative longitudinal magnetoresistivity, which is expected for a Weyl state. To access the predicted Weyl nodes, fine chemical potential tuning is probably necessary, which is beyond the scope of this work.

Besides magnetotransport measurements, we also performed laser-based ARPES measurements on one FM sample of MBST1 (15% Sb doping) to examine whether the non-trivial band topology is preserved in the MBST samples. The sample used for ARPES measurements was first screened by magnetization measurements and shows an FM transition nearly identical to those seen in other MBST1 samples (Fig. 3a). Figure 7 presents our ARPES measurement results. The electronic structure of the septuple layer (SL) terminated surface is displayed in Fig. 7a. The most important feature revealed in this spectrum is a Dirac-like band dispersion, with the Dirac point (DP) being at the binding energy of ~230 meV. Such a Dirac-like band dispersion can be attributed to the topological surface state (TSS). The observed TSS dispersion as well as its hybridization with a Rashba-like electron pocket seems similar to the ARPES results previously probed on the SL termination of AFM MnBi$_4$Te$_7$ [24,33,39,51]. Furthermore, we also performed circular dichroism (CD) ARPES measurements as shown in Fig. 7b. The CD in APRES is defined as the difference between the spectrum measured using left circularly polarized (LCP) light and the one measured using right circularly polarized (RCP) light, normalized by the sum of the two spectra. It is important to note that CD can be sensitive to spins as well as orbital angular momentum [52]. In MnBi$_2$Te$_4$(Bi$_2$Te$_3$)$_m$ systems, the CD pattern near the DP is consistent with the helical spin texture [53,54]. As shown in Fig. 7b, the CD pattern for the Dirac cone on the SL termination of MBST1 samples is fully antisymmetric with respect to the DP, reflecting the spin texture as expected for a TSS. Moreover, the electronic structure on the QL termination of our FM MBST1 sample also displays a clearly discernable TSS, as shown in Fig. 7(c) and 7(d). However, we did not observe sizable magnetic exchange gaps in the TSS for either termination, possibly due to the proximity of the measurement temperature (11 K) to $T_C$ (13.5 K). Our observations of the TSS with the chiral CD patterns clearly demonstrate the non-trivial band topology in our Sb-doped, FM samples.

## IV. DISCUSSIONS

### A. Origin of FM transitions in MBST.

Given that the only difference in synthesis conditions between samples MBT1 and MBT2 is that MBT2 experienced the additional thermal cycling treatment between 400 °C and 585 °C, their remarkable differences in magnetic transitions imply that the additional thermal treatment can generate disorders which significantly impact the interlayer magnetic coupling. Previous structure studies [42-44,55-58] on MnBi$_2$Te$_4$ and MnSb$_2$Te$_4$ have revealed that the major disorders in these materials originate from Mn-Bi(Sb) mixing at the main Mn layers and antisite Mn occupation at the Bi(Sb) site. Considering these disorders, their actual compositions can be expressed as $Mn_{1-2x-y}(Bi/Sb)_{2x+y}[(Bi/Sb)_{1-x}Mn_x]_2Te_4$ where $x$ and $y$ represent the antisite mixing concentration and Bi/Sb concentration on the Mn site, respectively [58]. The $x$ values obtained from structural refinements and composition measurements vary in the 0.01-0.09 range for $MnBi_2Te_4$ [22,48,56,58,59] (0.13-0.17 for $MnSb_2Te_4$ [43,44,58]), while $y$ is in the 0-0.18 range for $MnBi_2Te_4$ (0-0.22 for $MnSb_2Te_4$). Neutron scattering experiment has demonstrated the antisite Mn plays a critical role in driving ferrimagnetism in $MnSb_2Te_4$ [43,44]. Recent high field magnetization measurements, combined with theoretical simulations, demonstrate low density antisite Mn ions (~3.8%) can also drive ferrimagnetism in $MnBi_2Te_4$ [58]. This explains why the saturation moment of $MnBi_2Te_4$ (~ 4 $\mu_B$/f.u. at 10 T) is smaller than the expected value of 4.5 $\mu_B$ -5 $\mu_B$ [58].

Defects due to Mn-Bi mixing should also exist in $MnBi_4Te_7$ and other members of $MnBi_2Te_4(Bi_2Te_3)_m$. Experiments have shown that in $MnBi_4Te_7$ there is 18%-28% Bi on the Mn site and 1%-8% Mn at Bi sites [22,28,29,41]. Since the structure of $MnBi_4Te_7$ is composed of stacking of $[MnBi_2Te_4]$-SLs and $[Bi_2Te_3]$-QLs. Antisite Mn can occupy the Bi sites on both SLs and QLs. As indicated above, antisite Mn in SLs should be antiferromagnetically coupled with the main Mn layers, forming a ferrimagnetic-like SL block. Hu *et al.* [29] proposed the antisite Mn ions on the QLs can mediate FM stacking of ferrimagnetic SL blocks when they form a spin glassy state within the QL blocks below $T_N$. This model provides good explanations for the AFM-to-FM transitions observed in $Mn(Bi_{1-x}Sb_x)_4Te7$ [29]. Our observation of different magnetic properties between MBT1 and MBT2 as well as the FM transitions in samples MBST1-4 can also be well understood based on this model. The additional heat treatment experienced by MBT2 should result in more antisite Mn ions at the Bi sites of the SLs and QLs which not only enhances ferrimagnetic order within SLs but also increases interlayer FM coupling. As such, the interlayer magnetic coupling should involve AFM and FM competitions, thus resulting in magnetic glassy behavior at higher temperatures. This explains why MBT2 has higher $T_{ir}$ than MBT1. In the Sb-doped samples MBST1-4, the Sb doping should further increase the antisite Mn density since the ionic radius of $Sb^{3+}$ is closer to that of $Mn^{2+}$ than $Bi^{3+}$ [29]. The increased antisite Mn in QLs leads the FM stacking of ferrimagnetic SL blocks to have lower energy than the AFM stacking, thus resulting in a FM-like transition. This is not only consistent with the previous PBE-based DFT calculations, which show that the energy difference between interlayer AFM and FM coupling is as small as 0.23 meV for $MnBi_4Te_7$ [36], but also supported by our observations that the averaged saturation moments of samples MBST1-4 are 3-10% less than that of MBT1 (see Table 2). To further understand the magnetic phase stabilities of our synthesized materials, we also performed first-principles calculations based on density functional theory. We used a recently modified version of SCAN [60] ($r^2$SCAN[61]) with improved numerical stability by design. The state-of-art D4 dispersion correction method [62-64] was combined with $r^2$SCAN for a better description of van der Waals interactions. We calculated the energy differences between the FM and AFM phases ($E_{AFM}$-$E_{FM}$ per formula unit) of both pristine $MnBi_4Te_7$ and $(Mn_{0.75}Bi_{0.25})(Bi_{0.625}Sb_{0.325})_4Te_7$ considering (Mn,Bi) and (Bi,Sb) mixing. We find the AFM phase is more stable for the pristine phase; $E_{AFM}$ - $E_{FM}$ is -0.71 meV (-0.73 meV) with (without) spin-orbit coupling, while the FM phase is stabilized by defects for $(Mn_{0.75}Bi_{0.25})(Bi_{0.625}Sb_{0.325})_4Te_7$ with $E_{AFM}$ - $E_{FM}$ being 3.15 meV without spin-orbit coupling, which provides direct support for the above interpretation for the FM transition induced by Sb doping. Note that for the doped case, we didn't include the spin-orbit coupling due to the high computational cost. We expect the spin-orbit coupling doesn't affect the FM and AFM energy difference as in the pristine case. The reason why we could obtain FM $Mn(Bi_{1-x}Sb_x)_4Te_7$ with low Sb concentrations ($x = 0.15, 0.175, 0.2$ & $0.27$) is that we adopted a melt growth method with much shorter annealing time at 585 °C and a water quenching process as described above, which seems more effective in creating higher antisite Mn density than the methods reported in literature [21,22,28].

## B. Origin of magnetic glassy behavior

The magnetic relaxation data presented in Fig. 4 show both AFM MBT and FM MBST samples are characterized by magnetic glassy behavior at temperatures below $T_{ir}$. As noted above, Hu *et al*. [29] also observed magnetic glassy behavior in FM $Mn(Bi_{1-x}Sb_x)_4Te_7$ ($x = 0.76$), $Mn(Bi_{1-x}Sb_x)_6Te_{10}$ ($x = 0.07$) and $MnBi_8Te_{13}$ and attribute it to irreversible FM domain movement. Our observation of magnetic glassy behavior in both AFM and FM samples suggests domain wall movement is not a sole possibility. As shown in Fig. 2b & 2e and Fig. 3b, 3e & 3h, the AFM samples exhibit much stronger magnetic hysteresis than the FM samples, indicating that the FM domains driven by the spin-flip/flop transition in the AFM sample are more strongly pinned than the spontaneous FM domains in the FM samples. The similar slow magnetic relaxation seen in the AFM and FM samples suggests domain movement should not be the major cause of magnetic glassy behavior in our samples. Furthermore, we note the AFM $MnBi_4Te_7$ sample reported by Wu *et al*. [23] does not display magnetic glassy behavior, indicating that magnetic glassy behavior is sample dependent, and disorders should play a critical role in generating magnetic glassy behavior. In general, a magnetic system with disorders and magnetic frustration is expected to lead to spin glass or cluster spin glass states. The above discussions have shown that both $MnBi_4Te_7$ and Sb-doped $MnBi_4Te_7$ samples involve disorders and magnetic frustration. Their disorders are mostly caused by Mn-(Bi,Sb) mixing, as discussed above, and the magnetic frustration is driven the competing interlayer AFM and FM interactions, as evidenced by the small energy difference between these two magnetic states. With these facts, it is reasonable to interpret our observed magnetic glassy behavior in the AFM $MnBi_4Te_7$ or FM $Mn(Bi_{1-x}Sb_x)_4Te_7$ samples as cluster spin glass. However, all our samples exhibit AFM/FM transition signatures consistent with long-range magnetic orders below $T_N$ or $T_C$ and previous neutron scattering studies have demonstrated a long-range AFM order in $MnBi_4Te_7$ [21,22,29,42,44]. Therefore, the most reasonable scenario is that long-range AFM/FM orders coexist with AFM/FM cluster spin glass in our samples. Such coexistence of long-range magnetic orders with magnetic glassy behavior has been observed in other systems, such as $Co_3Sn_2S_2$ and PrAlGe [65,66].

### C. Possible topological phases in FM $Mn(Bi_{1-x}Sb_x)_4Te_7$

$MnBi_4Te_7$ can host a variety of topological phases. Bulk AFM $MnBi_4Te_7$ has been established as topological insulator and its non-trivial topological surface states have been detected in ARPES measurements by several groups [22,24,30,33,67,68]. Theory predicts that several other topological states, including quantum anomalous Hall insulator and quantum spin Hall insulator, can also be realized in 2D thin layers of AFM $MnBi_4Te_7$ via controlling the stacking pattern of SLs and QLs [31]. When the magnetic order in $MnBi_4Te_7$ becomes FM, it can support other forms of topological states. The theoretically predicted topological phases for FM $MnBi_4Te_7$ include an axion insulator, a topological crystalline insulator tunable by the magnetization orientation, and a Weyl semimetal as noted above [28,35]. However, MnBi$_4$Te$_7$ showing only a FM transition has never been reported. Although some reported MnBi$_4$Te$_7$ [29,37], as well as our MBT1&2 samples, exhibit AFM-to-FM transitions, our discussions presented above have shown such AFM-to-FM transitions are not uniform magnetic transitions, and the magnetic ground states most likely feature a mixture of AFM and FM phases. The other problem is that the AFM-to-FM transition happens below 5 K, which poses a challenge for probing the predicted topological phases using ARPES. As we have presented above, the FM phase of MnBi$_4$Te$_7$ is accessible via Sb doping to the Bi site. However, if the Sb doping concentration is too high ($x > 0.5$), the system may not preserve the non-trivial band topology [49]. Our ARPES measurements presented above (Fig. 7) has demonstrated that the non-trivial band topology is preserved in our FM MBST samples, consistent with a prior report which shows non-trivial band topology of the FM MnBi$_4$Te$_7$ sample doped by 30% Sb [37] [Note that Ref. [37] shows only 30% Sb-doped MnBi$_4$Te$_7$ is FM, while other samples with different concentrations of Sb (15%, 36-48%) show AFM-to-FM transitions]. This implies all of our MBST samples should possess a non-trivial band topology. The ferromagnetism combined with non-trivial band topology in such materials provide a platform to explore those predicted topological phases which occur in presence of ferromagnetism.

Like pristine MnBi$_4$Te$_7$ [22], our current FM samples are still electron-doped as manifested in Hall resistivity measurements (Fig. 6a, 6d & 6g). To verify the predicted axion insulator and the crystalline topological insulator, we need to make fine-tuning of Sb concentrations to tune the chemical potential into the gap. To observe the Weyl state, the FM sample needs to be lightly electron-doped since the calculated Weyl nodes are close to the bottom of the conduction bands. Finely tuning the chemical potential in FM $Mn(Bi_{1-x}Sb_x)_4Te_7$ is beyond the scope of the current work but the focus of our future work.

## V. CONCLUSIONS

In summary, we have gained a systematic control in the synthesis of FM $Mn(Bi_{1-x}Sb_x)_4Te_7$ with low Sb doping concentrations. Since low Sb concentration doping preserves non-trivial band topology, its combination with the time-reversal symmetry breaking caused by the spontaneous ferromagnetism is expected to realize those topological quantum phases predicted for FM $(MnBi_2Te_4)(Bi_2Te_3)_m$, including axion insulators, Weyl semimetals, quantum spin Hall insulators and thickness-independent quantum anomalous insulators. Furthermore, we have observed similar magnetic relaxation behavior in both AFM $MnBi_4Te_7$ and FM $Mn(Bi_{1-x}Sb_x)_4Te_7$ though the magnetic hysteresis of the former is much stronger than the latter, suggesting that the magnetic relaxation should originate from the long-range AFM/FM order coexisting with the cluster spin-glass phase. The combination of disorders and magnetic frustration arising from competing interlayer AFM and FM interactions should account for the cluster spin-glass phase. Finally, our analyses also suggest the antisite Mn plays a critical role in stabilizing the interlayer FM coupling.

**Acknowledgement**

This work was primarily supported by the US Department of Energy under grants DE-SC0019068 and DE-SC0014208. W.X. is supported by Beckman Young Investigator Award. S.H.L. is supported by the National Science Foundation through the Penn State 2D Crystal

**Table 1.** Single crystal XRD structure refinement for sample MBST1 at 296 K.

| Refined Formula | $(Mn_{0.80(1)}Bi_{0.20(1)})(Bi_{2.71(4)}Sb_{1.29(4)})Te_7$ |
|---|---|
| F.W. (g/mol) | 1700.25 |
| Space group; Z | $P\,3$; 1 |
| $a$ (Å) | 4.3436 (6) |
| $c$ (Å) | 23.801 (3) |
| V (Å$^3$) | 388.9 (1) |
| Extinction Coefficient | 0.003 (1) |
| θ range (deg) | 9.457-36.354 |
| No. reflections; $R_{int}$ | 7993; 0.0863 |
| No. independent reflections | 2428 |
| No. parameters | 43 |
| $R_1$: $\omega R_2$ ($I>2\delta(I)$) | 0.0517; 0.1329 |
| Goodness of fit | 1.096 |
| Diffraction peak and hole (e$^-$/ Å$^3$) | 7.174; -5.652 |
| Absolute structure parameter | 0.49 (2) |

Table 2: Sample information of $Mn(Bi_{1-x}Sb_x)_4Te_7$, including magnetic ordering temperature, saturation moment at 2K, carrier density & mobility at 200K, and anomalous Hall resistivity $\Delta\rho_{xy}$ at 2 K (or 2.5 K). Since the saturation moment is sample dependent even for the samples from the same batch (which is possibly caused by chemical inhomogeneity), we present variation ranges as well as averaged values ($\bar{M}_S$) for each batch. For $\Delta\rho_{xy}$, we also present two values measured on two different samples for each batch except MBST2.

| Sample Label | Sb Content, $x$ (Nominal Value) | Magnetic Ordering ($T_N$ or $T_C$) | Saturation Moment ($\bar{M}_s$) ($\mu_B$/f.u.) | Carrier Density ($10^{20}$ cm$^{-3}$) | Mobility (cm$^2$/Vs) | $\Delta\rho_{xy}$ ($\mu\Omega$ cm) |
|---|---|---|---|---|---|---|
| *MBT1* | 0 | AFM (13 K) | 2.55 - 3.01 (2.82) | 1.8 | 43.9 | 6.81, 9.17 |
| *MBT2* | 0 | AFM (13 K) | 2.29 - 3.11 (2.70) | 2.2 | 69.1 | 5.33, 5.18 |
| *MBST1* | 0.15 | FM (13.5K) | 2.53 - 2.73 (2.66) | 0.65 | 188 | 14.1, 7.88 |
| *MBST2* | 0.175 | FM (13.5K) | 2.52 - 2.60 (2.54) | 0.87 | 184 | 5.03 |
| *MBST3* | 0.2 | FM (13.5K) | 2.55 - 3.05 (2.74) | 0.28 | 315 | 9.26, 9.48 |
| *MBST4* | 0.27 | FM (13K) | 2.40 − 2.77 (2.60) | 0.70 | 124 | 8.14, 5.41 |

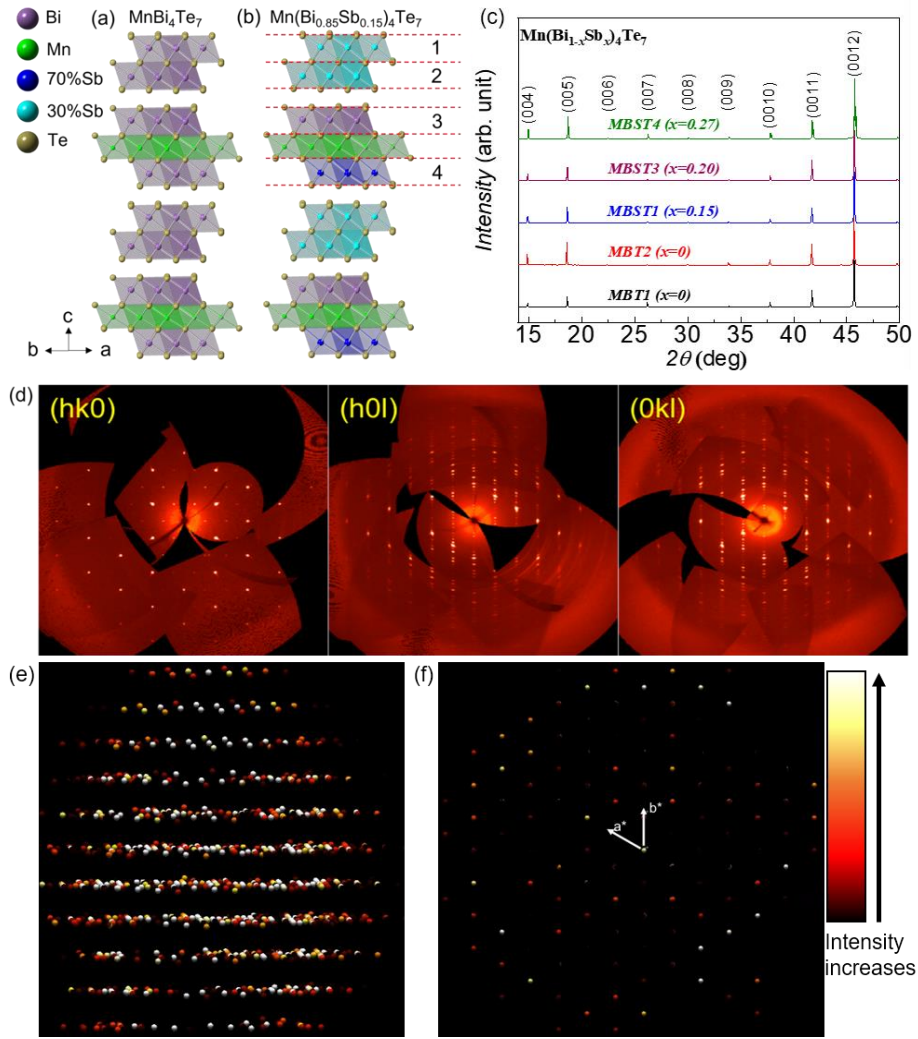

**Fig. 1**. Schematic crystal structures of pristine $MnBi_4Te_7$ (a) and $Mn(Bi_{0.85}Sb_{0.15})_4Te_7$ (b). $Mn(Bi_{0.85}Sb_{0.15})_4Te_7$ shows preferential Sb occupations at different Bi sites on layers labeled by 1, 2, 3, and 4. (c) X-ray diffraction patterns of samples MBT1, MBT2, MBST1, MBST3, and MBST4. (d) Single-crystal X-ray diffraction patterns obtained on the ($hk$0), ($h$0$l$), and (0$kl$) planes. (e) Reflection intensity map in the reciprocal space, projected along a zone-axis slightly away from $c^*$. (f) The intensity map along the $c^*$-axis.

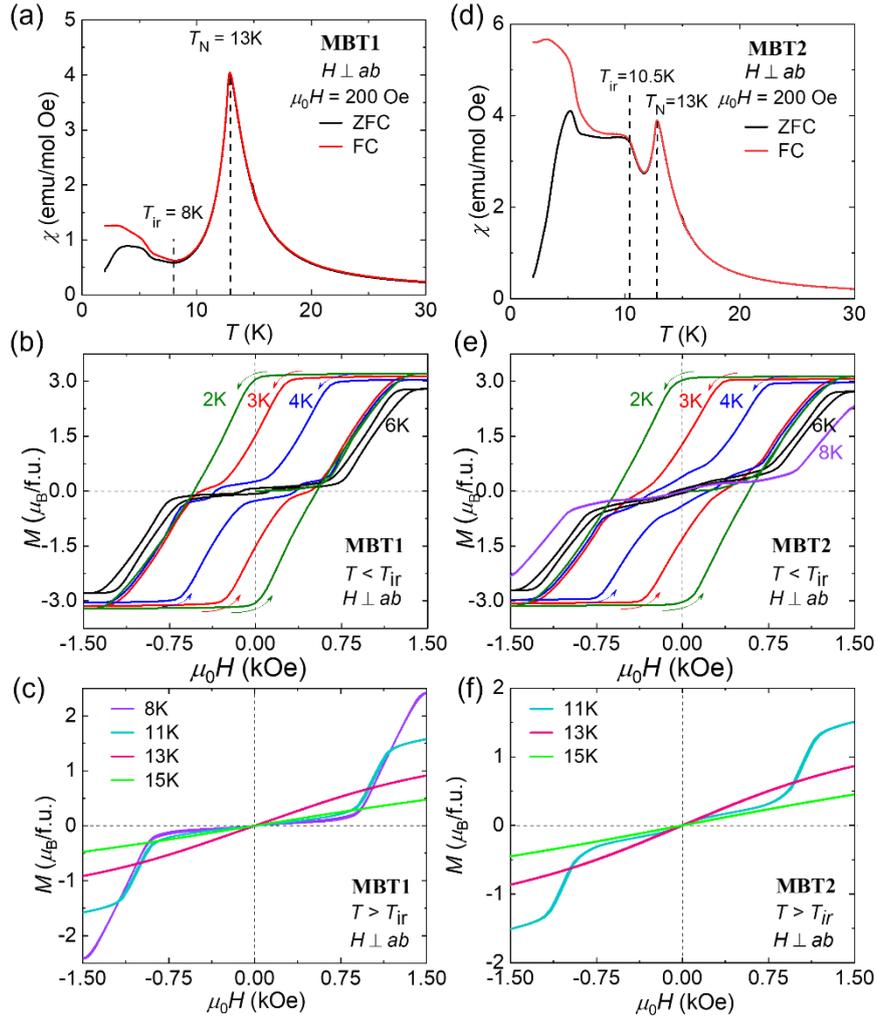

**Fig. 2.** (a,d) Temperature dependence of magnetic susceptibility measured under ZFC and FC histories for (a) MBT1 and (d) MBT2. Isothermal magnetization measured with $H \perp ab$-plane for (b,c) MBT1 and (e,f) MBT2 at temperatures below $T_{ir}$ (b,e) and above $T_{ir}$ (c,f).

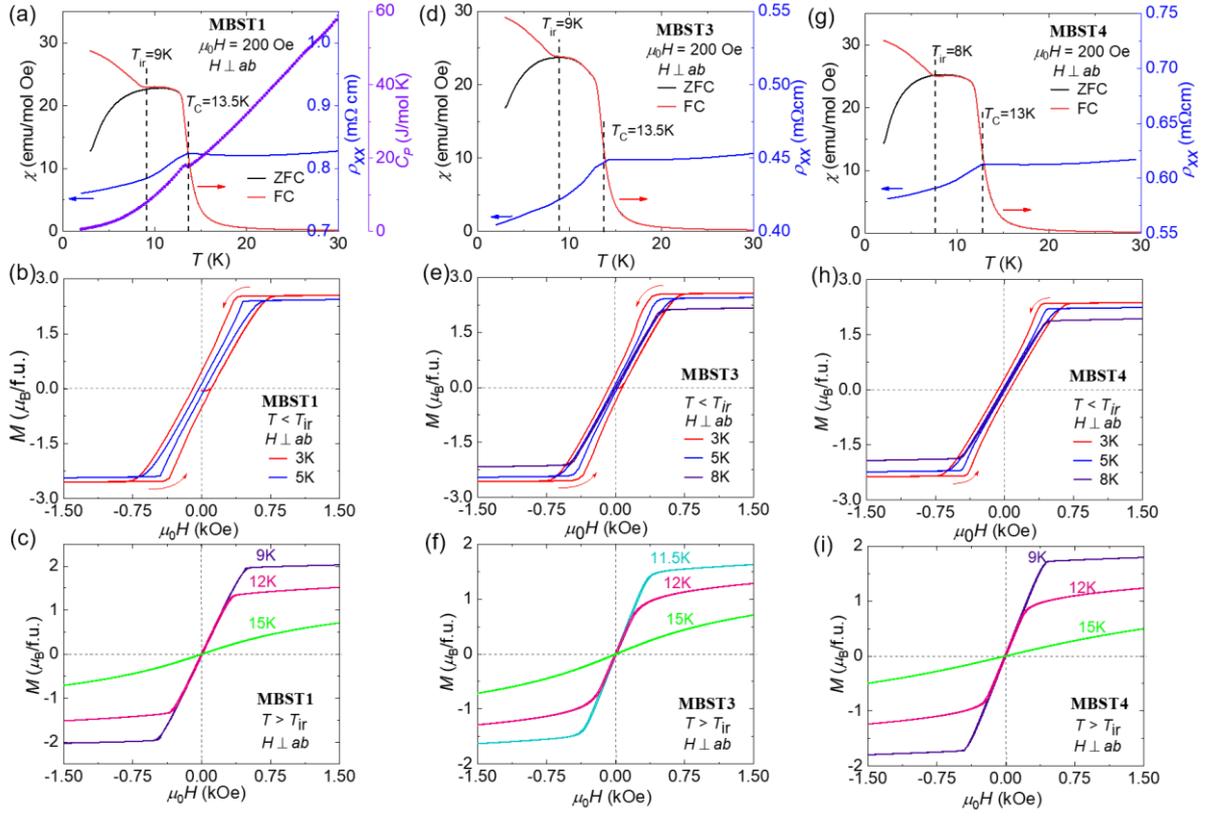

**Fig. 3.** (a, d, g) Temperature dependence of magnetic susceptibility measured under ZFC and FC histories for (a) MBST1, (d) MBST3 and (g) MBST4. The blue and the purple curves in (a, d, g) represent the temperature dependence of in-plane resistivity $\rho_{xx}$ and specific heat for (a) MBST1, (d) MBST3 and (g) MBST4, respectively. Isothermal magnetization measured with $H \perp ab$-plane for (b, c) MBST1, (e, f) MBST3, and (h, i) MBST4 at temperatures below $T_{ir}$ (b, e, h) and above $T_{ir}$ (c, f, i).

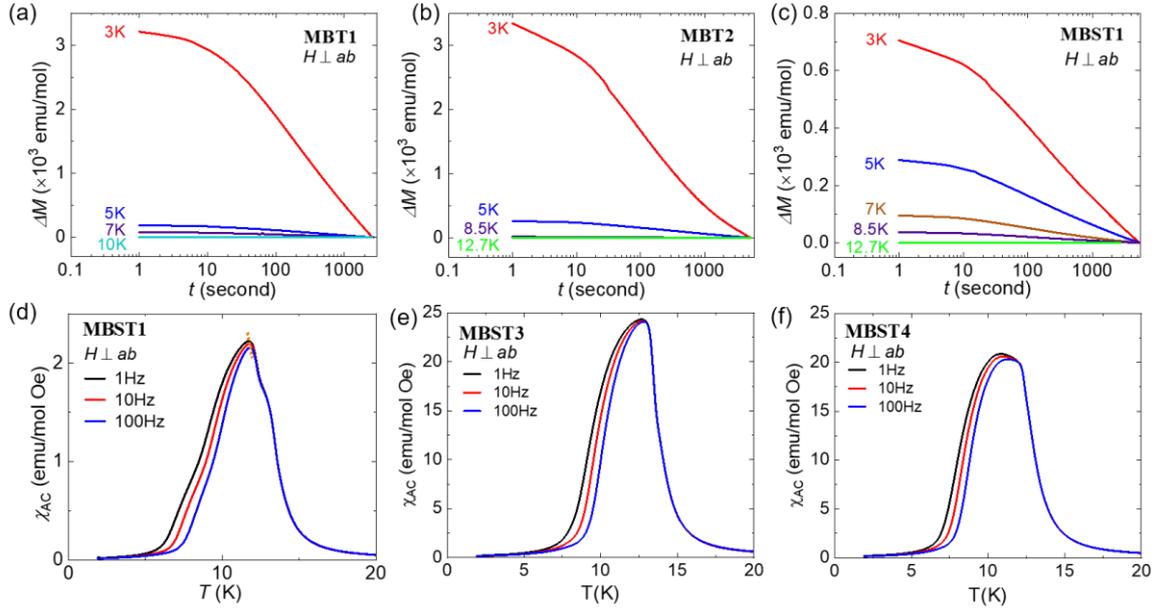

**Fig. 4.** Time dependence of magnetization measured right after ramping the magnetic field up to 5000 Oe ($H \perp ab$-plane) and back to zero at various fixed temperatures for (a) MBT1, (b) MBT2, and (c) MBST1. Temperature dependence of AC susceptibility at various frequencies for (d) MBST1, (e) MBST3, and (f) MBST4.

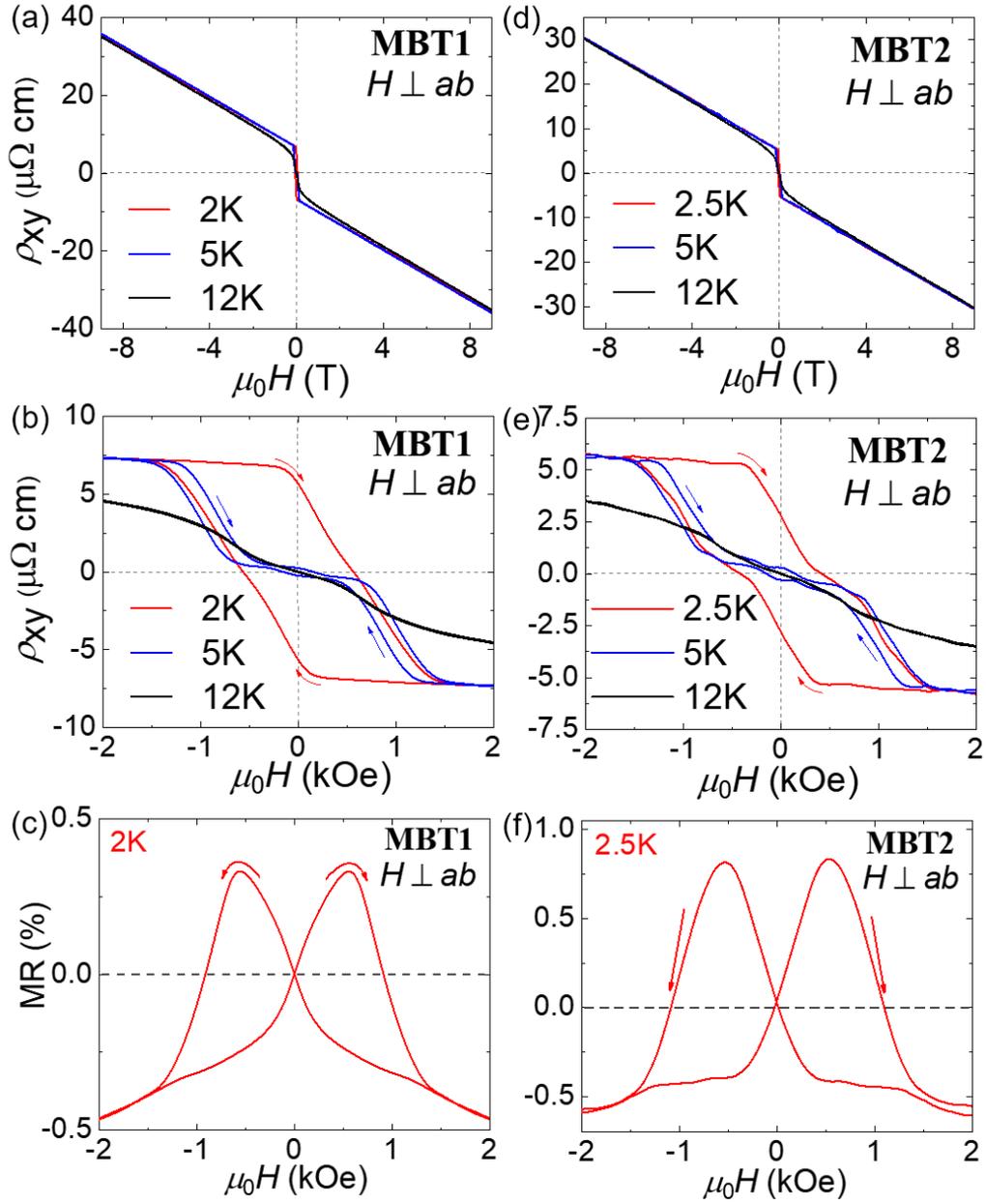

**Fig. 5.** (a, b, d, e) Field dependence of Hall resistivity $\rho_{xy}$ measured with $H \perp ab$-plane in the -9 T – 9 T (a, d) and -2 kOe – 2 kOe (b, e) field ranges at various temperatures for MBT1 (a, b) and MBT2 (d, e). Transverse magnetoresistivity MR measured with $H \perp ab$-plane for (c) MBT1 and (f) MBT2 at 2.0 K and 2.5 K.

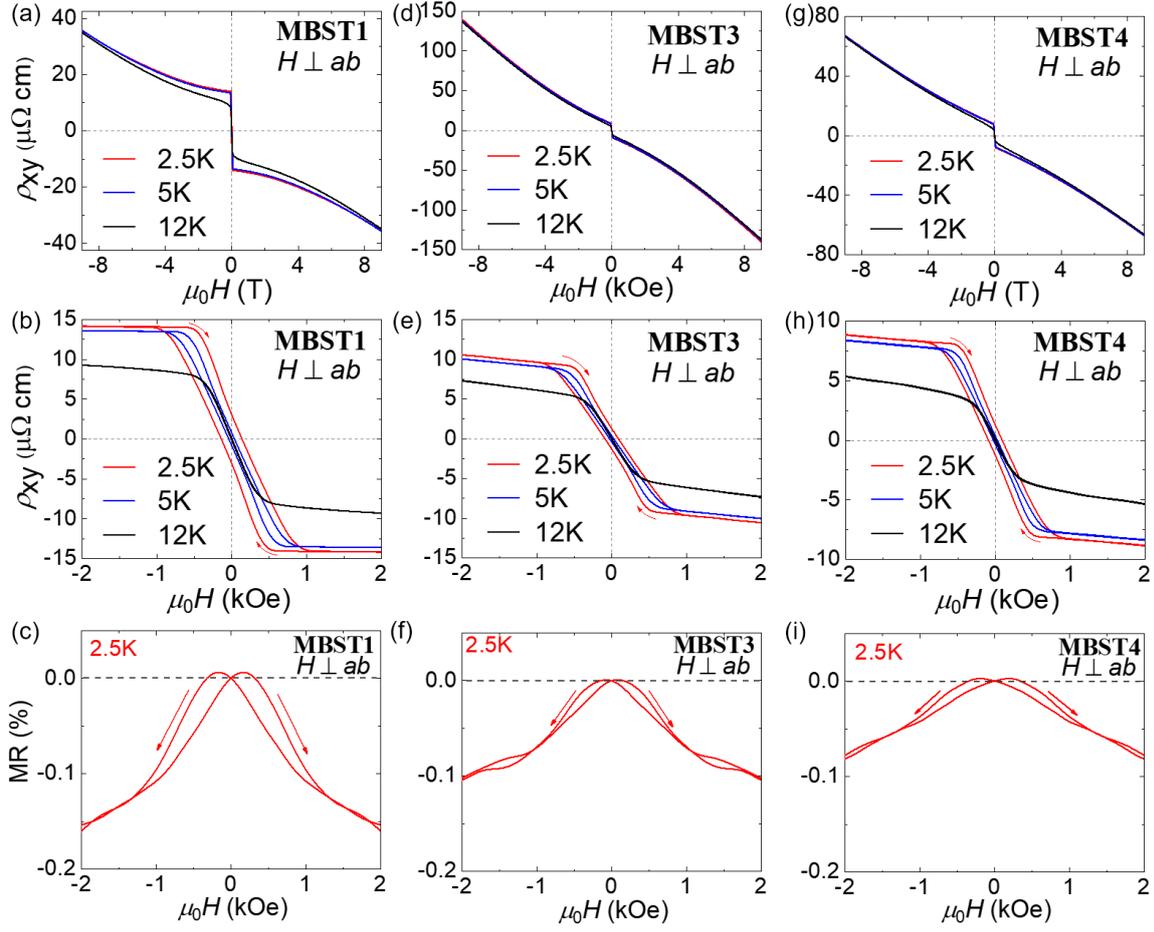

**Fig. 6.** (a, b, d, e, g, h) Field dependence of Hall resistivity $\rho_{xy}$ measured with $H \perp ab$-plane in the -9 T – 9 T (a, d, g) and -2 kOe – 2 kOe (b, e, h) field ranges at various temperatures for MBST1 (a, b), MBST3 (d, e), and MBST4 (g, h). Transverse magnetoresistivity MR measured with $H \perp ab$-plane for (c) MBST1, (f) MBST3 and (i) MBST4 at 2.5 K.

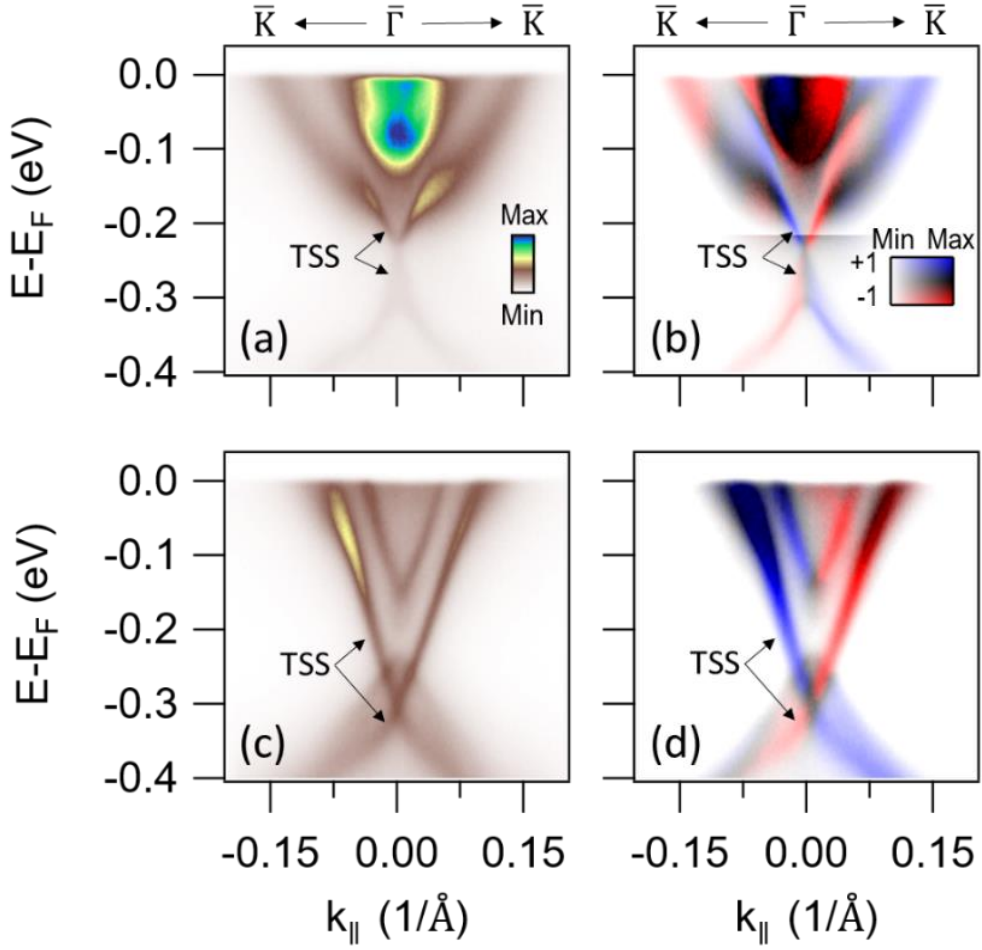

Fig 7. Electronic structures for two terminations of $Mn(Bi_{0.85}Sb_{0.15})_4Te_7$ (MBST1). (a) ARPES spectrum on the septuple layer (SL) termination along the $\overline{\Gamma} - \overline{K}$ direction at 11 K. (b) Circular dichroism (CD) spectrum for the SL termination. CD is obtained by taking the difference between the two spectra using left and right circularly polarized light, normalized by the sum. We adopt a 2D color scale where the red-blue contrast represents the CD pattern, and the overall saturation represents the spectral intensity. The intensity scale is adjusted for the lower branch of the Dirac cone to account for the much weaker spectral intensity. The counterpart results for the quintuple layer (QL) termination are plotted in (c) and (d).

**Supplementary Material**
**for**
**"Ferromagnetic MnBi$_4$Te$_7$ obtained with low concentration Sb doping: a promising platform for exploring topological quantum states"**

Y.D. Guan[1+], C.H. Yan[2+], S.H. Lee[1,3], X. Gui[4,] W. Ning[1], J.L. Ning[5], Y.L. Zhu[1,3], M. Kothakonda[5], C.Q. Xu[6], X.L. Ke[6], J.W. Sun[5], W.W. Xie[7], S.L. Yang* and Z.Q. Mao[1,3,8*]

[1] Department of Physics, The Pennsylvania State University, University Park, PA 16802, USA

[2] Pritzker School of Molecular Engineering, The University of Chicago, Chicago, Illinois 60637, USA

[3] 2D Crystal Consortium, Materials Research Institute, The Pennsylvania State University, University Park, PA 16802, USA

[4] Department of Chemistry, Louisiana State University, Baton Rouge LA 70803

[5] Department of Physics and Engineering Physics, Tulane University, New Orleans, LA 70118

[6] Department of Physics and Astronomy, Michigan State University, East Lansing, MI 48824, USA

[7] Department of Chemistry and Chemical Biology, Rutgers University, Piscataway NJ 08854

[8] Department of Materials Science and Engineering, The Pennsylvania State University, University Park, PA 16802, USA

**Supplementary Note 1**
We performed DFT calculations for the ground state total energy for the ferromagnetic phases in three defect scenarios, each with uniform and preferential Sb occupations, respectively, as shown in Supplementary Table S1. The composition of sample MBST2 (20% Sb) is close to scenario (iii). From the calculated total energy results, we find the energy of the FM state with uniform Sb occupation is ~ 4-15 meV/unit-cell lower than the FM state with preferential Sb occupation. Such a small difference might be within the uncertainty range of DFT calculations.

Although the DFT calculations cannot resolve the total energy difference of the ground state between uniform and preferential Sb occupations, lattice dynamics at finite temperature might be different for the uniform and preferential Sb occupations; preferential Sb occupations may have lower free energies at high temperatures in comparison with the uniform Sb occupations (Note that like $MnBi_2Te_4$, $MnBi_4Te_7$ is also a metastable phase and can be synthesized only through quenching at high temperatures). We do not perform DFT-based lattice dynamics calculations due to the prohibitively high computational cost associated with the large supercell size needed for the systems considered. Additionally, antisite defects also play a critical role in stabilizing the FM phase as discussed in the main text.

Table S1: The calculated ground state total energy for the ferromagnetic phases with uniform and preferential Sb occupations for three different scenarios of antisite defects and Sb substitution for Bi: (i) 37.5% Sb occupying Bi sites and 25% Bi occupying Mn sites; (ii) 37.5% Sb and 6.25% Mn occupying Bi sites and 25% Bi occupying Mn sites; (iii) 25%Sb and 6.25%Mn occupying Bi sites and 25% Bi occupying Mn site.

| Formula | Sb Occupation | FM energy (eV) |
| --- | --- | --- |
| $(Mn_3Bi)(Bi_{10}Sb_6)Te_{28}$ | Preferential | -1573.137 |
| $(Mn_3Bi)(Bi_{10}Sb_6)Te_{28}$ | Uniform | -1573.154 |
| $(Mn_3Bi)(Bi_9MnSb_6)Te_{28}$ | Preferential | -1530.218 |
| $(Mn_3Bi)(Bi_9MnSb_6)Te_{28}$ | Uniform | -1530.279 |
| $(Mn_3Bi)(Bi_{11}MnSb_4)Te_{28}$ | Preferential | -1596.542 |
| $(Mn_3Bi)(Bi_{11}MnSb_4)Te_{28}$ | Uniform | -1596.574 |

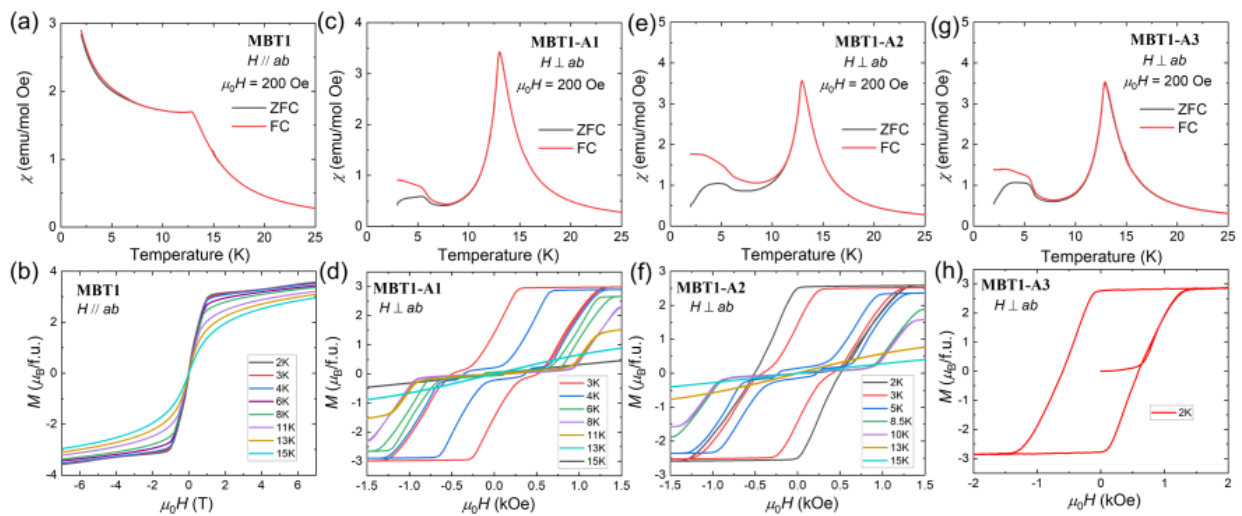

Figure S1: (a) Temperature dependence of magnetic susceptibility measured with $H//ab$ for the MBT1

sample shown in Fig. 2(a-c) in the main text. (b) Isothermal magnetization at various temperatures measured with *H//ab* for the MBT1 sample shown in Fig. 2(a-c). (c, e, g) Temperature dependences of magnetic susceptibility measured with $H \perp ab$ for three additional MBT1 samples (the second number in the sample label refers to piece number). (d, f, h) Isothermal magnetization at various temperatures measured with $H \perp ab$ for three additional MBT1 samples.

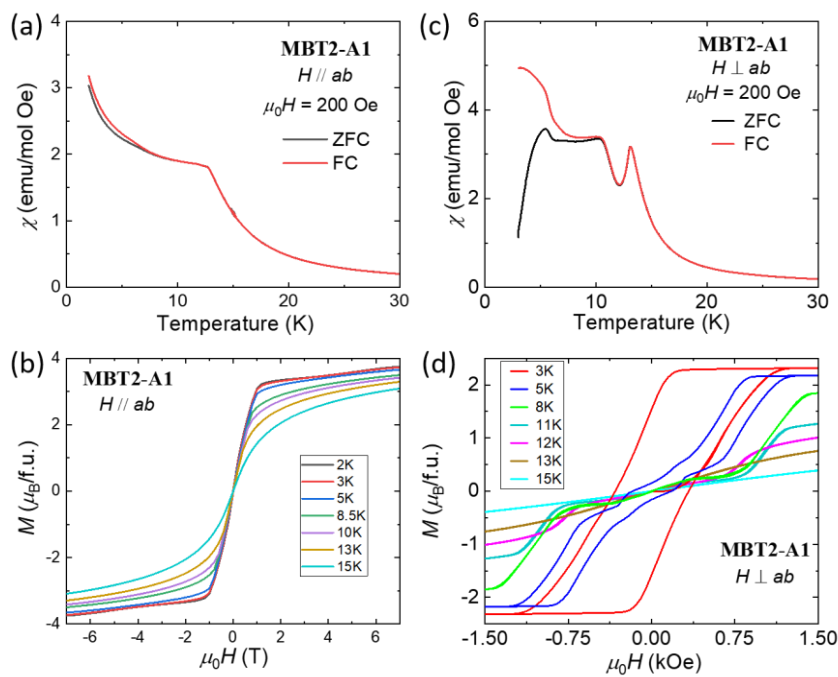

Figure S2: Temperature dependence of magnetic susceptibility measured with *H//ab* (a) and $H \perp ab$ (c) for an additional MBT2 sample (MBT2-A1). Isothermal magnetization at various temperatures measured with *H//ab* (b) and $H \perp ab$ (d) for sample MBT2-A1.

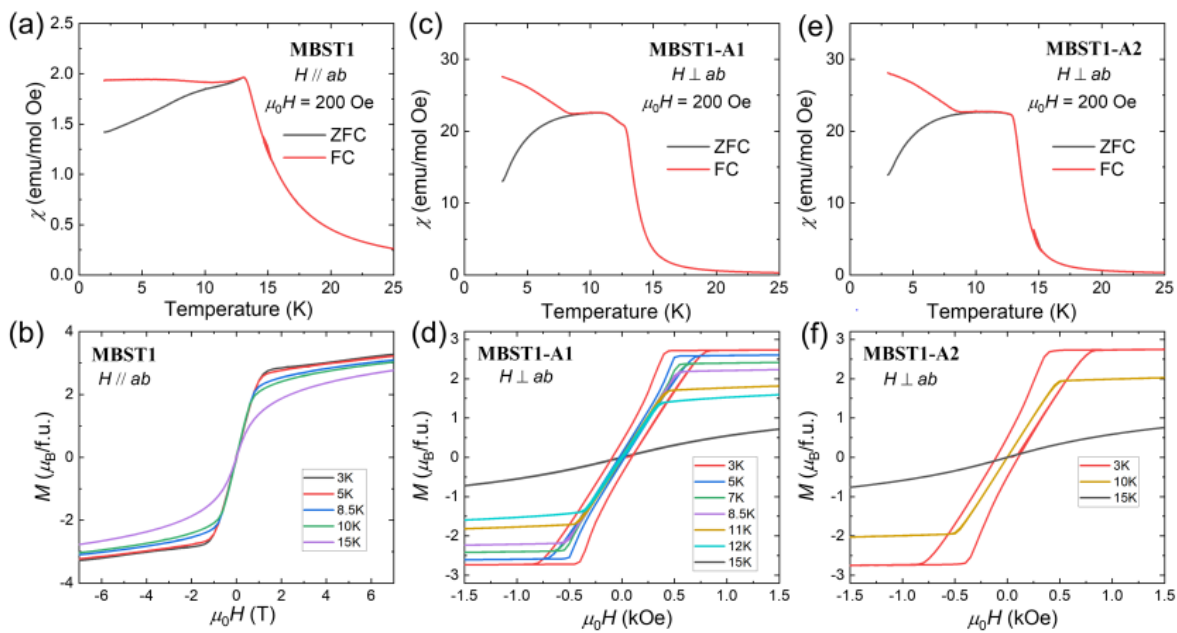

Figure S3: (a) Temperature dependence of magnetic susceptibility measured with *H//ab* for the MBST1

sample shown in Fig. 3(a-c) in the main text. (b) Isothermal magnetization at various temperatures measured with $H//ab$ for the MBST1 sample shown in Fig. 3(a-c). (c, e) Temperature dependences of magnetic susceptibility measured with $H \perp ab$ for two additional MBST1 samples (the second number in the sample label refers to piece number). (d, f) Isothermal magnetization at various temperatures measured with $H \perp ab$ for two additional MBST1 samples.

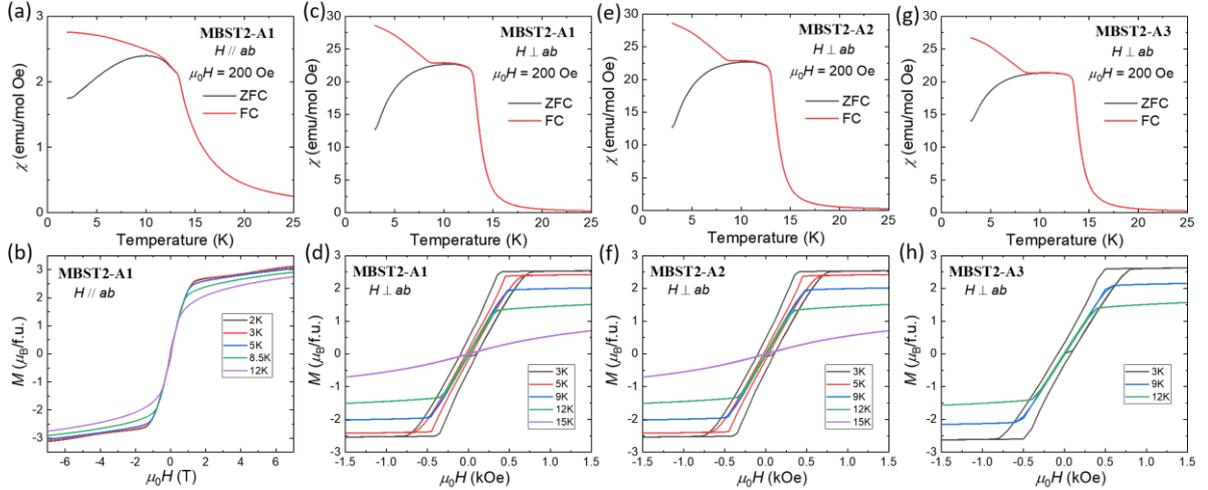

Figure S4: (a, c) Temperature dependence of magnetic susceptibility measured with $H//ab$ and $H \perp ab$ respectively for a MBST2 sample. (b, d) Isothermal magnetization at various temperatures measured with $H//ab$ and $H \perp ab$ for the MBST2 sample shown in Fig. S4(a, c). (e, g) Temperature dependences of magnetic susceptibility measured with $H \perp ab$ for another two additional MBST2 samples. (f, h) Isothermal magnetization at various temperatures measured with $H \perp ab$ for the additional MBST2 samples shown Fig. S4(e, g).

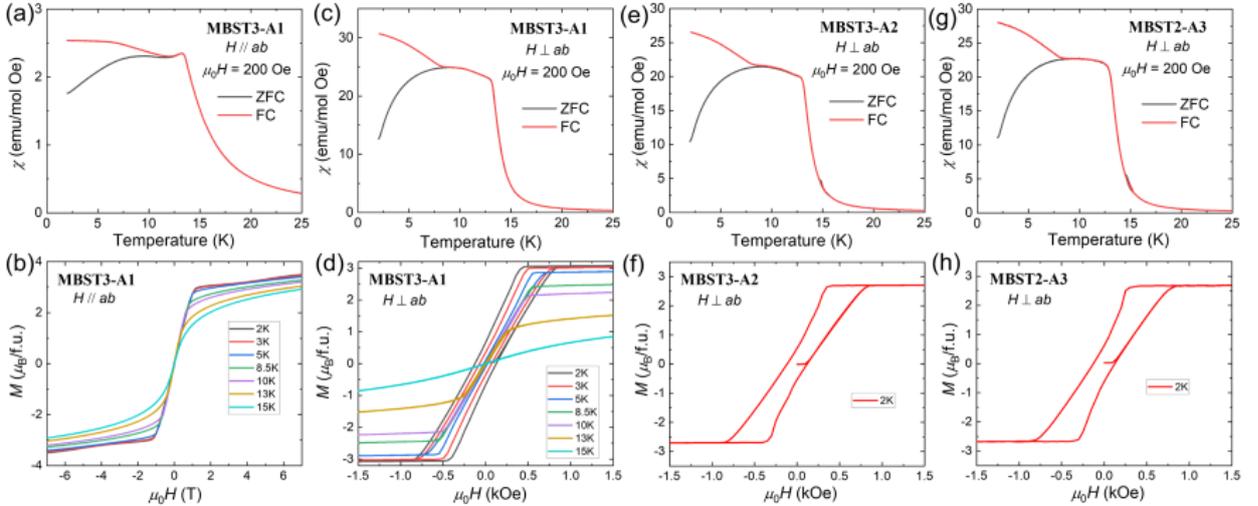

Figure S5: (a, c) Temperature dependence of magnetic susceptibility measured with $H//ab$ and $H \perp ab$ respectively for a MBST3 sample. (b, d) Isothermal magnetization at various temperatures measured with $H//ab$ and $H \perp ab$ for the MBST3 sample shown Fig. S5(a, c). (e, g) Temperature dependences of magnetic susceptibility measured with $H \perp ab$ for another two additional MBST3 samples. (f, h) Isothermal magnetization at 2 K measured with $H \perp ab$ for the additional MBST3 samples shown Fig. S5(e, g).

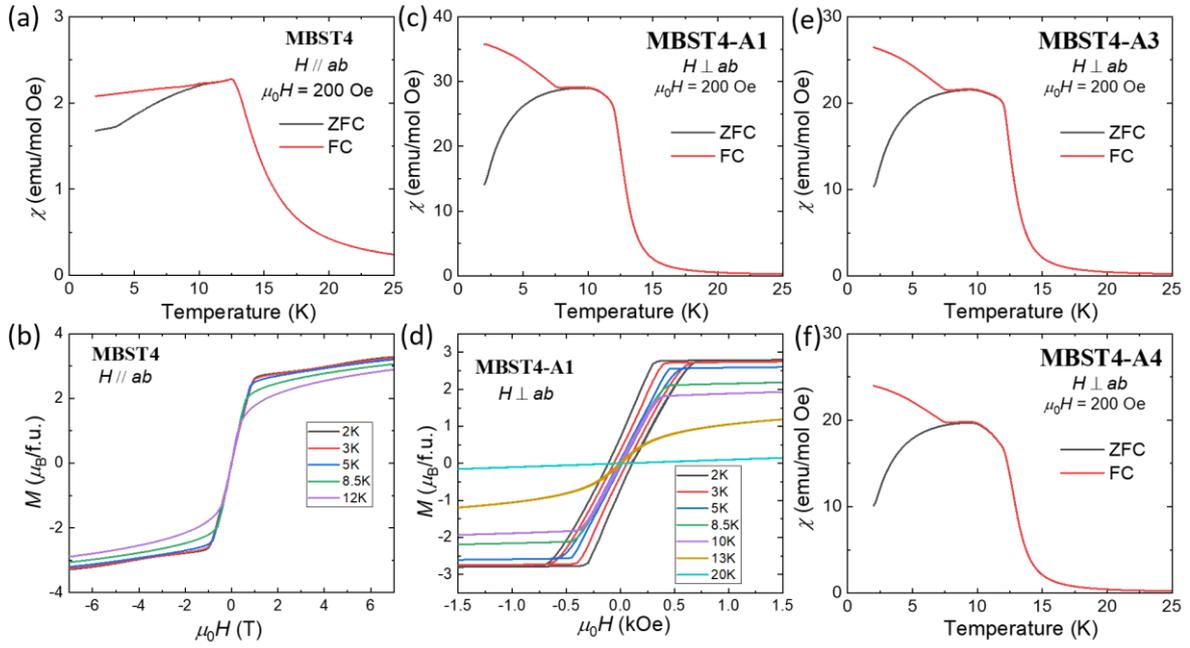

Figure S6: (a) Temperature dependence of magnetic susceptibility measured with $H//ab$ for the MBST4 sample shown in Fig. 3(g-i) in the main text. (b) Isothermal magnetization at various temperatures measured with $H//ab$ for the MBST4 sample shown in Fig. 3(g-i). (c, e, f) Temperature dependence of magnetic susceptibility measured with $H \perp ab$ for three additional MBST4 samples. (d) Isothermal magnetization at various temperatures measured with $H \perp ab$ for the sample shown in Fig. S6(c) (MBST4-A1).

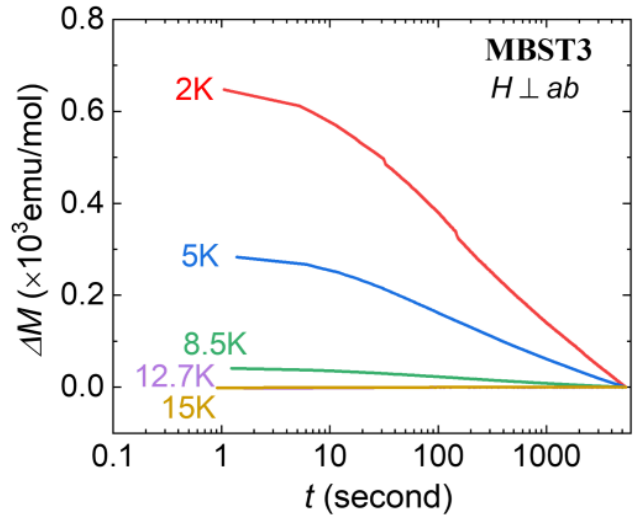

Figure S7: Time dependences of magnetization measured right after ramping the magnetic field up to 5000 Oe ($H \perp ab$-plane) and back to zero at various fixed temperatures for sample MBST3. $\Delta M = M(t) - M_0$ where $M_0$ represents the magnetization at 5,000 s, the time when the measurement ends.

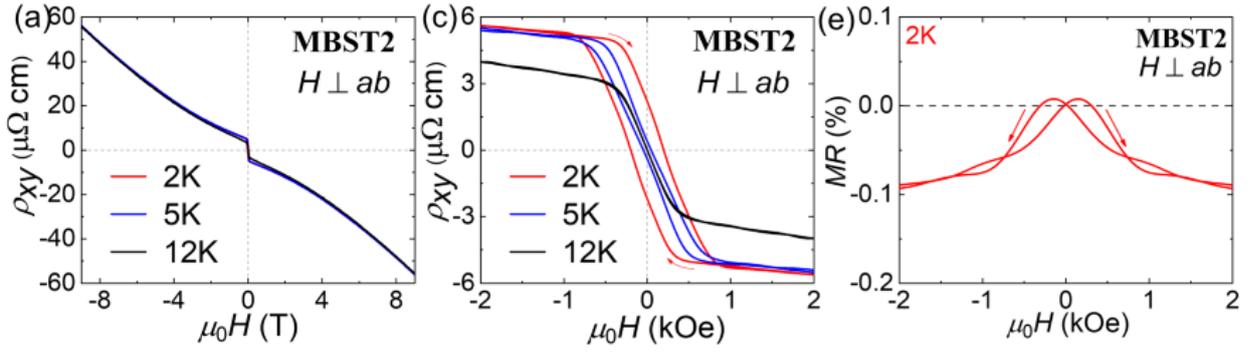

Figure S8: Hall resistivity as a function of a magnetic field at 2 K, 5 K, and 12 K in the -9 T – 9 T field range (a), the -0.2 T – 0.2 T field range (b) for sample MBST2. (c) In-plane transverse magnetoresistance at 2 K for MBST2.

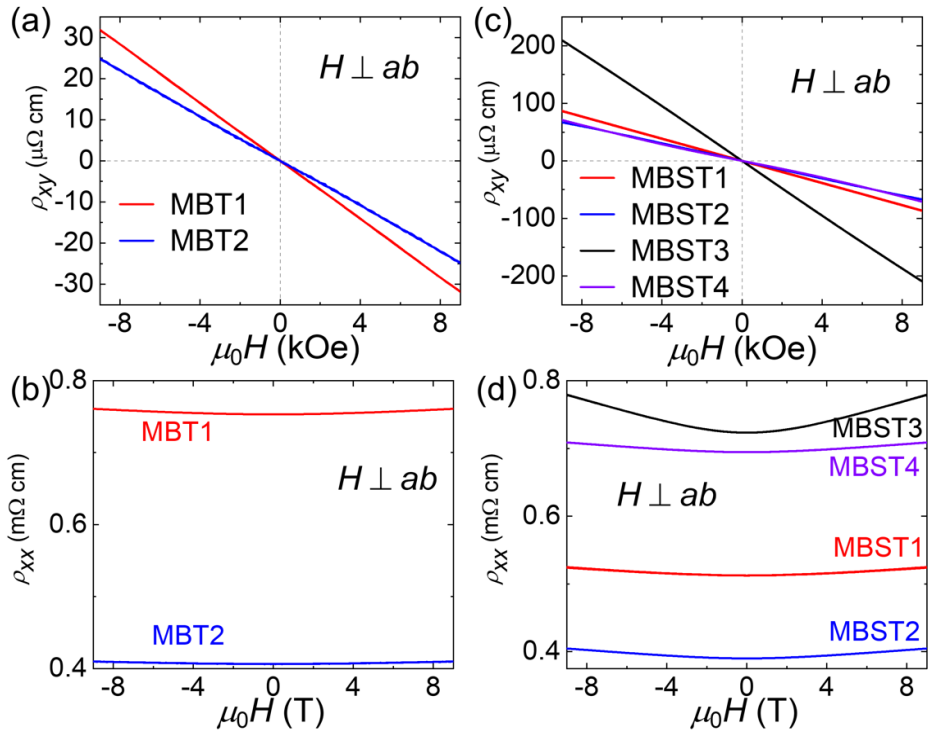

Figure S9: Hall resistivity as a function of magnetic field at 200 K for samples MBT1&2 (a), MBST1-4 (c). (b, d) Field dependence of in-plane transverse magnetoresistivity at 200 K for MBT1&2 and MBST1-4.

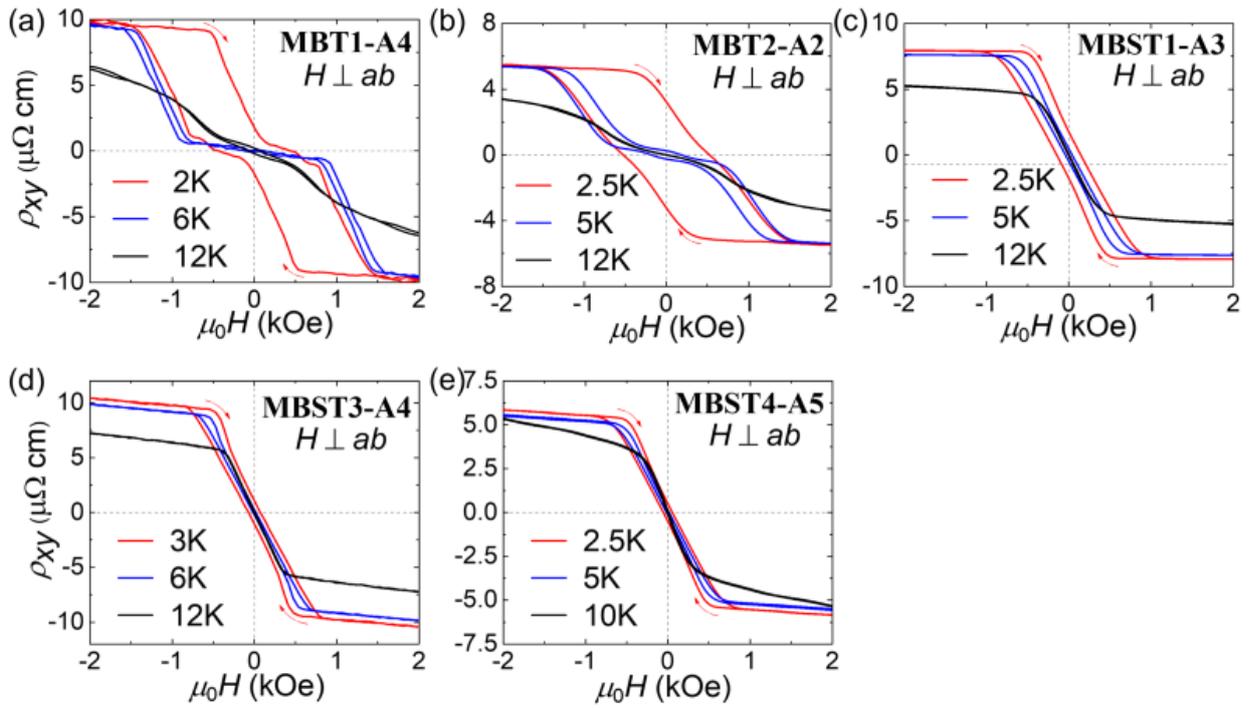

Figure S10: Hall resistivity as a function of magnetic field at various temperatures in the -0.2 T – 0.2 T field range for additional MBT1, MBT2, MBST1, MBST3, and MBST4 samples.

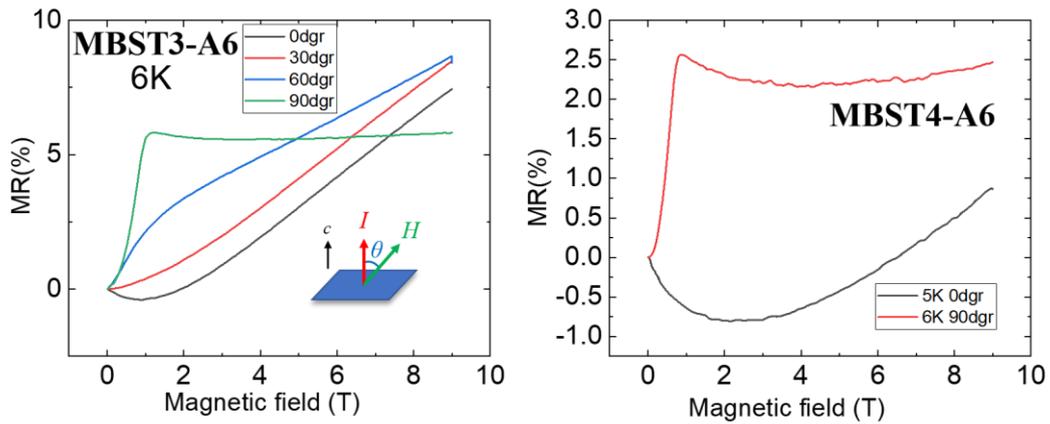

Figure S11: Field dependence of in-plane magnetoresistivity under various field orientations at 6 K (6 K/ 5 K for MBST-A6) for samples MBST3 (left) and MBST4 (right).